\documentclass[amsmath,amssymb,aps,prl,tightenlines,twocolumn]{revtex4-1}

\usepackage[english]{babel}
\usepackage{afterpage}
\usepackage{graphicx}
\usepackage{dcolumn}
\usepackage{bm}
\usepackage{color}
\usepackage{slashed}
\usepackage{latexsym}

\begin{document}

\title{Laser-induced electron Fresnel diffraction by XUV pulses at extreme intensity}

\author{Lei Geng$^1$}
\author{Hao Liang$^1$}
\author{K. Krajewska$^2$}
\author{Liang-You Peng$^{1,3,4,5}$}
\email{liangyou.peng@pku.edu.cn}
\author{Qihuang Gong$^{1,3,4,5}$}

\affiliation{
$^1$State Key Laboratory for Mesoscopic Physics and Frontiers Science Center for Nano-optoelectronics, School of Physics, Peking University, 100871
Beijing, China\\
$^2$Institute of Theoretical Physics, Faculty of Physics, University of Warsaw, Pasteura 5, 02-093 Warsaw, Poland \\
$^3$Collaborative Innovation Center of Quantum Matter, Beijing 100871, China\\
$^4$Collaborative Innovation Center of Extreme Optics, Shanxi University, 030006 Taiyuan, China \\
$^5$Beijing Academy of Quantum Information Sciences, Beijing 100193, China
}

\begin{abstract}
 Ionization of atoms and molecules in laser fields can lead to various interesting interference structures in the photoelectron spectrum. For the case
 of a super-intense extreme ultraviolet laser pulse, we identify a novel petal-like interference structure in the electron momentum distribution along
 the direction of the laser field propagation. We show that this structure is quite general and can be attributed to the Fresnel diffraction of
 the electronic wavepacket  by the nucleus. Our results are demonstrated  by numerically  solving the  time-dependent Schr\"odinger  equation of
 the atomic hydrogen  beyond the dipole approximation. By building an analytical model, we find that  the electron displacement determines the
 aforementioned interference pattern. In addition, we establish the physical picture of laser-induced electron Fresnel diffraction which is reinforced by both quantum and semiclassical models.
\end{abstract}
\maketitle

 As a fundamental law of quantum mechanics, the superposition of coherent electronic wavefunctions will result in interference. When exposed to external laser fields, the electrons in atoms and molecules can be ionized at different temporal and spatial positions~\cite{Arbo2010,Xu_2011,Waitz2016}, which leads to various interference structures in the photoelectron spectrum. The dynamical and structural information~\cite{Gopal2009,Korneev2012,Arbo2006} is encoded in the spectrum through the phase of electronic wavefunction~\cite{Figueira_de_Morisson_Faria_2020}. Experimentally, one can acquire the information about the targets by measuring
the photoelectron momentum distribution~(PMD) using the cold-target recoil-ion-momentum spectroscopy~(COLTRIMS)~\cite{Ullrich1997} or the velocity map
imaging~(VMI) spectrometers~\cite{Chandler87,Eppink97}. The theoretical analysis can be carried out by numerical solution of the time-dependent
Schr\"odinger equation~(TDSE) with high precision~\cite{Huismans2011,Jiang2020}, or by more intuitive semiclassical methods~\cite{Li2014, Geng2015, Shvetsov2016}
through the integration of the Newton equation of the electron, together with  its classical phase~\cite{Figueira_de_Morisson_Faria_2020}.

While the works mentioned above were carried out within the dipole approximation, strong-field physics beyond that approximation has attracted increasingly more interest in recent years~\cite{Wang_2020}. It has been shown both experimentally  and theoretically that nondipole corrections to the Hamiltonian can induce a photon-momentum transfer to the photoelectron~\cite{Smeenk2011,Chelkowski2014,He2017,Wang2017,Liang2018,Chen2020,Ni2020}. Besides, when atoms are exposed to the extreme ultraviolet (XUV) pulse of high intensity, the
nondipole corrections can alter the dynamic interference~\cite{Wang2018}.
With the further increase of the laser intensity, the nondipole effects become even more remarkable and the photoelectron angular distribution can be largely modified~\cite{Forre2006}
due to the strong interplay between the external electromagnetic and the internal Coulomb forces. The short-wavelength lasers at extreme intensities
have been an everlasting pursuit~\cite{Young_2018}. In this regime, many new  phenomena will be further identified.

In this Letter, we theoretically investigate  the ionization dynamics of atoms exposed to a super-intense XUV laser pulse, based on the accurate numerical
solution to TDSE beyond the dipole approximation. Our TDSE method is based on the finite element discrete variable representation~\cite{McCurdy2000} and
the Arnoldi propagator~\cite{Park1986}. In the low-energy part of the PMD, we observe a novel petal-like interference structure along the direction of
the pulse propagation. In this regime of laser parameters, the structure is quite  robust and can be attributed to the  Fresnel diffraction of the electronic wavepacket by the nucleus. Quantum and semiclassical models have been proposed to quantitatively reproduce the petal-like structures, which corroborate the physical mechanism of the laser-induced electron Fresnel diffraction.

Quantum-mechanical dynamics of an electron in a classical electromagnetic field is governed by the Schr\"odinger equation $i\partial_t\Psi({\mathbf r},t)=H\Psi({\mathbf r},t)$, in which  the nondipole Hamiltonian $H$ in the propagation gauge and the long-wavelength approximation can be written as~\cite{Forre2016,Forre2020}
\begin{align}
    H&= \frac{{\mathbf p}^2}{2}+{\mathbf A}(t)\cdot{\mathbf p}+V({\mathbf r})+\frac{{\mathbf A}^2(t)({\hat {\mathbf k}}\cdot{\mathbf p})}{2c}\nonumber\\
    & +\frac{({\hat {\mathbf k}}\cdot{\mathbf p})({\mathbf A}(t)\cdot{\mathbf p})}{c}-\frac{({\hat {\mathbf k}}\cdot{\mathbf r})({\mathbf \nabla}V({\mathbf r})\cdot{\mathbf A}(t))}{c},
\label{Hamiltonian}
\end{align}
where all corrections of order $1/c$  beyond the dipole approximation have been included. (Atomic units are employed throughout the Letter unless otherwise
stated, where $c\approx137\,{\rm a.u.}$ is the speed of light.) Note that, in the dipole approximation, the Hamiltonian only contains the first three terms
on the right hand side of Eq.~(\ref{Hamiltonian}).

In this Letter, we consider a linearly polarized laser pulse with the propagation vector $\hat{\mathbf k}$ in the $z$-direction.
Its vector potential ${\mathbf A}(t) $ is given by
\begin{equation}
{\mathbf A}(t)=\frac{\sqrt{I_0}}{\omega} \exp(-\frac{4\ln2\, t^2}{\tau^2}) \sin(\omega t) {\hat{\mathbf e}_x},
\label{laser}
\end{equation}
where $I_0$ is the peak intensity,  $\omega$  is the carrier frequency, and  $\tau$  represents the full width at half maximum~(FWHM).
The corresponding electric field  is ${\mathbf E}(t)=-\mathrm{d}{\mathbf A}(t)/\mathrm{d} t$.
\begin{figure}
   \includegraphics[width=\linewidth]{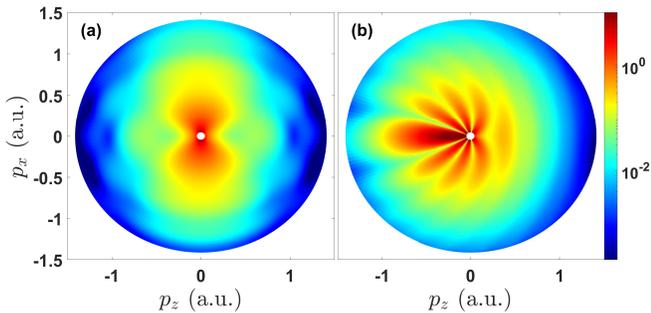}
  \caption{The PMDs in the $xz$-plane calculated either within~(a) or  beyond~(b)  the dipole approximation. See the main text for laser parameters.
  }
  \label{PMD}
\end{figure}

Let us first examine the ionization of the ground state of the hydrogen atom by a laser pulse with $\omega=3\,{\rm a.u.}$,  $\tau=2\sqrt{2}\pi\, {\rm a.u.}$,
and $I_0=4\times10^{20}\,{\rm W/{cm}^2}$. In Fig.~\ref{PMD}, we show the PMDs within and beyond the dipole approximation. Only the low-energy parts
below the one-photon ionization peak are demonstrated as the majority of the ionization occurs in this region. Although the spectra in both cases center around the zero momentum,  they exhibit very distinct features.  The PMD calculated within the dipole approximation is elongated in the $x$-direction. In the nondipole case, on the other hand, the PMD is most pronounced along the negative $z$-axis and is asymmetric about the $x$-axis. Compared to the simulations in  Ref.~\cite{Forre2006}, novel petal-like interference  lobes are present.
We have numerically checked that the first nondipole term ${\mathbf A}^2(t)({\hat {\mathbf k}}\cdot{\mathbf p})/(2c)$ in Eq.~\eqref{Hamiltonian}, which originates from
the radiation pressure~\cite{Reiss2014,Smeenk2011}, is mainly responsible for the nondipole effects considered in our case. This term also indicates that a magnetic force ${\mathbf F}_{\rm m}=-{\rm A}(t){\rm E}(t)/c {\hat{\mathbf e}_z}$ is imposed on the electron~\cite{Forre2006}.

\begin{figure}
     \includegraphics[width=\linewidth]{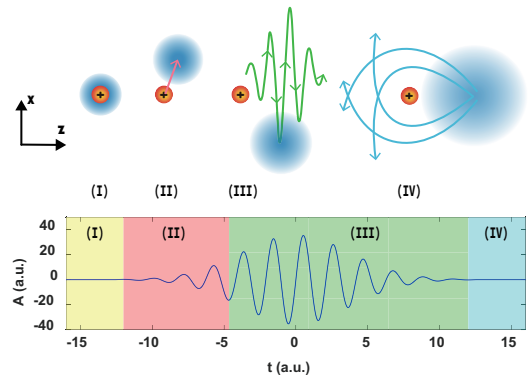}
    \caption{The cartoon representing the ionization  process via four  steps. The lines with arrows represent the trajectories of electrons. (I) The hydrogen atom is stationary without the laser field.
    (II) The electronic wave packet is non-adiabatically pulled away by the pulse in the rising portion of the envelope.
    (III) Next, it oscillates with the vector potential and moves forwards in the direction of pulse propagation.
    (IV) Finally, the electronic wave packet is diffracted by the nucleus. In addition, after step (II), the wave packet diffuses with time.
    Here, the nucleus is fixed at each step of the process whereas the vector potential as a function of time is plotted at the bottom.
    }
    \label{sketch}
\end{figure}

In order to get more insight into the dynamics of photoelectrons, we present a movie in the Supplemental Materials~\cite{movie}, illustrating how
the probability density distribution in the $xz$-plane varies in time.  We note that the entire electronic wave packet is first pulled away from the nucleus.
This is in contrast to the atomic stabilization in the Kramers-Henneberger frame~\cite{eberly1993}, where the wave packet is split into two portions along the
direction of field polarization and stays stable against ionization.  This difference follows the fact that the adiabatic condition is not satisfied in our case~\cite{dorr2000}. After being  pulled away, the electronic wave packet drifts in the laser field in the $\hat{\mathbf k}$ direction due to the magnetic force ${\mathbf F}_{\rm m}$. After the end of the laser pulse, the group velocity of the electronic wave packet
is close to zero. Then, the diffusion of the wave packet induces spherical waves centered at a fixed position. The nucleus can be deemed
as an obstruction for the induced spherical waves. One can compare this situation to the Fresnel diffraction  in wave optics where the distance from the source to the obstruction is finite.

\begin{figure*}
  \centering
   \includegraphics[width=0.8\linewidth]{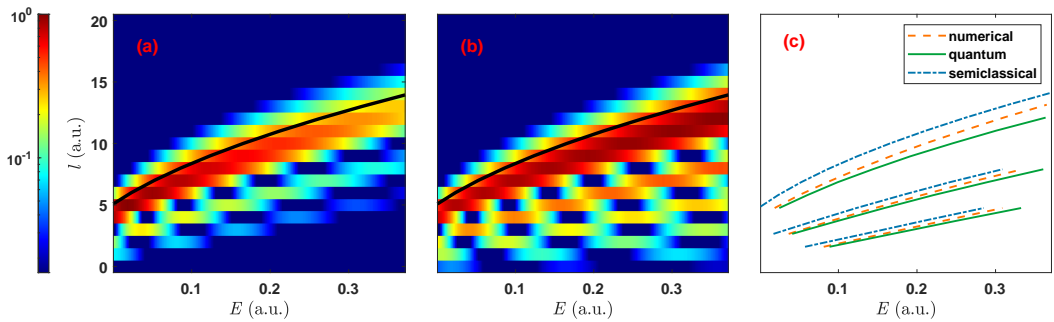}
  \caption{Color mappings of the energy-angular momentum distribution calculated from the TDSE~(a) and quantum model~(b), together with the relation~\eqref{classical}
  (black line). The comparison of the peak positions of these results is shown in (c), against the the semiclassical predictions~\eqref{other_branches}  for $n=0,1,2$ from the top to the bottom, respectively.
  }
  \label{E_L}
\end{figure*}

The above mechanism can be depicted by a cartoon shown in Fig.~\ref{sketch}.
During the quiver motion of the electron, we have seen the isotropic diffusion of its wave packet~\cite{movie}, which indicates that the Coulomb force
does not affect much the electron dynamics much when the intense laser field is present. Neglecting the Coulomb interaction, we obtain that the electron has a velocity component
${A^2}(t)/(2c){\hat{\mathbf e}_z}$, which is induced by the magnetic force. Hence, a classical electron displacement $z_0$ after the pulse~\eqref{laser}
can be calculated,
\begin{equation}
 z_0=\int_{-\infty}^{\infty}\frac{A^2(t)}{2c} dt\approx\sqrt{\frac{\pi}{2{\rm ln}2}}\, \frac{I_0 \tau}{8\omega^2c}.
 \label{displacement}
\end{equation}
 In wave optics, the distance from the source to the obstruction determines the Fresnel diffraction. Similarly, we shall show now
 that our diffraction patterns mainly depend  on the classical displacement~\eqref{displacement}. This will be demonstrated analytically and
confirmed numerically for various pulse durations and carrier frequencies.

According to Refs.~\cite{Arbo2006(2), Arbo2008}, the photoelectron angular distribution can be decomposed into contributing partial waves.
Thus,  one can examine the energy-angular momentum distributions of the ionized electrons.
It was shown in Refs.~\cite{Arbo2006(2),Arbo2008} that, for a low-energy electron ionized by an infrared laser, its angular momentum satisfies a classical relation
 $l=\sqrt{2\alpha+2\alpha^2 E}$, where $E$ is the photoelectron energy and $\alpha$ is the amplitude of its quiver motion. For the low-energy spectrum in our case, we can carry out similar decompositions and  the normalized
$(E,l)$ joint distributions are presented in Fig.~\ref{E_L}(a), in which the black line  shows the special case of $\alpha=z_0$, i.e.,
\begin{equation}
  l=\sqrt{2z_0+2z_0^2 E}.
  \label{classical}
\end{equation}
 It is intriguing that the black line coincides with the main branch of the distribution though the electron is ionized by a very different pulse.
 Note that the classical angular momentum has been decreased by 0.5 for a better comparison with the TDSE results. Besides the main one, branches with smaller $l$ also appear in Fig.~\ref{E_L}(a).  They lead to the asymmetry of the PMDs about the $x$-axis.
 To further interpret our {\it ab initio} results in terms of the Fresnel diffraction, we recall that the electron dynamics can be described by a set
 of classical trajectories including phase in the semiclassical picture~\cite{Li2014,Shvetsov2016}. In our case, the induced spherical waves imply
 classical electrons with the same probability of emission in all directions. They are launched at $z_0$ when the pulse ends, which promises that there
 is more than one classical trajectory corresponding to each $(E,l)$ state,  some of which have been intuitively depicted in Fig.~\ref{sketch}~(IV).
 For a given angular momentum, there are two trajectories with the same energy that will contribute to interference (for electrons moving towards
 and away from the nucleus). The branches correspond to the interference enhancement of contributing orbits in the semiclassical picture.
 It can be shown that the $(E,l)$ relation is determined by~\cite{semiclassical}
\begin{equation}
    \int_{r_{\text{min}}}^{z_0}\sqrt{2(E-\frac{l^2}{2r^2}+\frac{1}{r})}dr=n\pi,
  \label{other_branches}
\end{equation}
 where $r_{\text{min}} =  {(\sqrt{1+2El^2}-1)}/{2E}$ and $n$ is a non-negative integer. These semiclassical results for $n=0,1,2$ are plotted in Fig.~\ref{E_L}(c). The case for $n=0$ follows Eq.~\eqref{classical}, which corresponds to the condition that the initial velocity of the electron is perpendicular to the $z$-axis in the semiclassical picture.

Our {\it ab initio} numerical  results can also be understood based on a simplified quantum model. It is known that the radial part of the  scattering
state  $R_{kl}(r)$ is analytically given in terms of  the confluent hypergeometric function~\cite{Landau1981}. For the present case,  the electronic wave function
after the end of the pulse can be approximately written as $\delta(r-z_0)\delta(\cos\theta-1)/r^2$, the $(E,l)$ distribution can then be shown to be
\begin{equation}
  \left| a(E,l)\right|^2\propto\frac{2l+1}{\sqrt{E}}\left|R_{\sqrt{2E}l}(z_0)\right|^2.
  \label{a_l_e}
\end{equation}
The results based on this formula are shown for comparison in Fig.~\ref{E_L}(b), which reproduces the main interference features in Fig.~\ref{E_L}(a).
There do exist some differences, e.g.,  the simple quantum model overestimates the signals for smaller $l$ components. One also notes that the discrepancies become larger with the  increase of  $E$. They may be attributed to the finite  volume of the electronic wave packet and to the effect of the Coulomb potential on the quiver motion of the electron.

For a better comparison, in Fig.~\ref{E_L}(c), we  draw the peak positions of results from the {\it ab initio} calculations and the quantum model against the  prediction of the semiclassical model. Indeed, this comparison confirms that the numerically identified petal-like interference structures can be reproduced quantitatively by the simple quantum model and be understood qualitatively by the semiclassical one, which are based on the  electron Fresnel diffraction by the nucleus.

\begin{figure}
   \includegraphics[width=\linewidth]{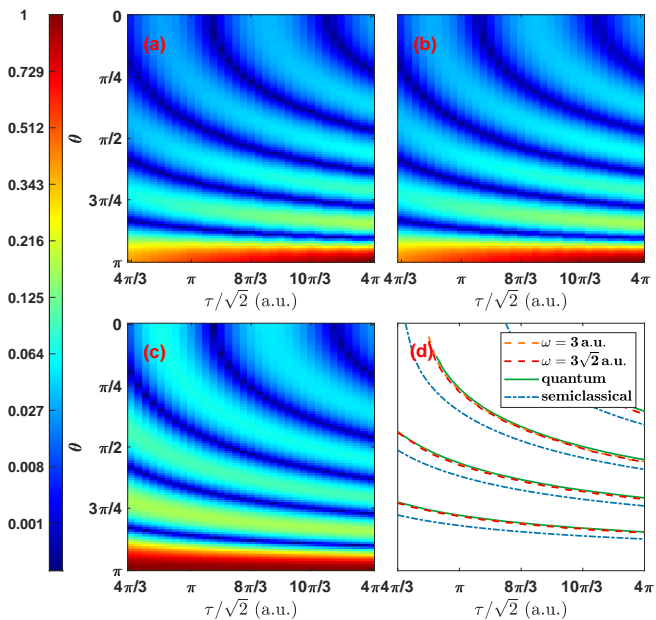}
  \caption{The normalized angular momentum distribution  at $E_s=10^{-4}\,{\rm a.u.}$ for different FWHM $\tau$.
  The numerical color mappings are for $\omega=3\,{\rm a.u.}$ (a), $\omega=3\sqrt{2}\,{\rm a.u.}$ (b), and the  analytical result (c) [Eq.~\eqref{a_quantum}]. The comparison of the peak positions of different results is shown in (d).
  }
  \label{T_theta}
\end{figure}

As can be seen from Eqs.~\eqref{other_branches}~and~\eqref{a_l_e}, the $(E,l)$ distributions of photoelectrons are determined by the classical displacement $z_0$. This in turn indicates that, in both semiclassical and quantum models, $z_0$ plays a major role in
forming the diffraction patterns in PMDs. As it follows from the definition of the displacement $z_0$ [Eq.~\eqref{displacement}], it changes with the peak intensity, the carrier frequency, and
the duration of the laser pulse. To confirm the general validity of our models based on  the  Fresnel diffraction, numerical calculations of  two other cases have been carried out. In the first case, the FWHM of the laser pulse $\tau$
varies from $4\sqrt{2}\pi/3\,{\rm a.u.}$ to $4\sqrt{2}\pi\,{\rm a.u.}$, whereas other parameters of the pulse are kept the same as before. In the second case,
the duration $\tau$ is also gradually varied in the same interval, but the carrier frequency and the peak intensity of the pulse is changed to $\sqrt{2}\omega$  and  $2I_0$ respectively. Note that, for both cases, the displacement $z_0$ is the same for
 a given $\tau$, which means that one should obtain the same PMDs.  Without loss of  generality for the discussion of the diffraction, one can consider the case of a small energy at $E_s=10^{-4}\,{\rm a.u.}$    In Figs.~\ref{T_theta}(a) and~\ref{T_theta}(b),  we show the normalized  photoelectron angular distributions at $E_s$ for the  two cases respectively.
First, we find that both color mappings are identical, which meets our expectations. In addition, we observe that
the number of peaks in the angular distribution of photoelectrons increases for a longer pulse duration, which can  be successfully accounted for
by our  semiclassical and quantum models.  For the quantum model, when $E\rightarrow 0$, the scattering state of Coulomb potential becomes
$R_{kl}/\sqrt{k}\, |_{\lim_{k \to 0}}=\sqrt{4\pi/r}J_{2l+1}(\sqrt{8r})$, where $J$ is the Bessel function~\cite{Landau1981}. Therefore, the angular distribution can be written as
\begin{equation}
  \left| a(\theta,z_0) \right|^2\propto \left| \sum_{l=0}^{l_{\max}} \frac{2l+1}{\sqrt{z_0}}J_{2l+1}(\sqrt{8z_0})P_l(\cos\theta)\right|^2,
  \label{a_quantum}
\end{equation}
 where $P_l$ is the Legendre polynomial and $l_{\max}$ is set to 25 in the present calculations. The normalized result according to Eq.~\eqref{a_quantum} is presented in Fig.~\ref{T_theta}(c),  which
 quantitatively reproduces the patterns shown in Figs.~\ref{T_theta}(a) and (b).
 Equivalently,  the angular distribution patterns can also be qualitatively interpreted by our semiclassical model. For the zero-energy case, the semiclassical orbits become
 parabolas and thus the angular distribution can be described analytically as~\cite{semiclassical}
 \begin{equation}
    \left| a(\theta,z_0) \right|^2\propto \left| 1+\exp[i4\sqrt{2z_0}\sin( {\theta}/{2})]\right|^2.
  \label{a_semiclassical}
\end{equation}
It can be proven that the quantum model leads to the same PMDs as those of the semiclassical model  when $z_0\to\infty$. For a better comparison,
the peak positions of results from both  TDSE and  the quantum model are drawn against that of the semiclassical model, as shown in Fig.~\ref{T_theta} (d). We can see that the  {\it ab initio} results for the  two cases almost coincide with the prediction of the quantum model, and the result of the semiclassical model qualitatively agrees with them.

Before concluding, we shall  stress the differences of the diffraction discussed here from the laser-induced electron
diffraction for molecules and rescattering for atoms, which are known in the literature~\cite{Yurchenko2004,Meckel2008,Becker2018}.
For the latter two cases, recolliding electrons can be represented by plane waves with nonzero effective momenta. This can be interpreted as the Fraunhofer
diffraction. In our case, however, the group velocity of the electron after it interacts with the pulse is zero, so the diffusion of the electron
induces spherical waves centered at the distance $z_0$. These are the spherical waves rather than plane waves that diffract at the Coulomb potential.
In other words, as already shown, we deal here with the Fresnel-type of diffraction. Though two types of diffraction are both induced by the combined
action of the laser pulse and Coulomb force, the Fresnel-type diffraction induces more low-energy photoelectrons and is usually slower
than the other one.

In conclusion, we have studied the dynamics of photoelectrons interacting with a super-intense XUV pulse. We found a novel petal-like diffraction
structure along the laser pulse propagation direction in the low-energy part of the PMDs, which have been attributed to the laser-induced Fresnel
diffraction of the electronic wave packet by the nucleus.  Based on this diffraction picture, an intuitive semiclassical and a simplified quantum model
have been developed and can analytically account for the novel interference patterns due to the nondipole effects. In addition, we show that this
diffraction phenomenon is quite robust and general in the regime of super-intense XUV pulses.

This work is supported by the National Natural Science Foundation of China~(NSFC) under Grant Nos.~11961131008 and 11725416, by the National Key R\&D Program of China under Grant No. 2018YFA0306302, and by the National Science Centre (Poland) under Grant No. 2018/30/Q/ST2/00236.

\nocite{landau1976mechanics}
\nocite{feynman2010quantum}
\bibliography{mybibtex}
\end{document}